\documentclass[prl,amssymb,twocolumn,superscriptaddress,showpacs]{revtex4}
\usepackage{bm}
\usepackage{graphicx}
\usepackage{color}
\usepackage{amsmath}

\begin{document}

\title{Observation of the underscreened Kondo effect in a molecular transistor}

\author{Nicolas Roch}
\affiliation{Institut N\'eel, associ\'e \`a l'UJF, CNRS, BP 166,
38042 Grenoble Cedex 9, France}
\author{Serge Florens}
\affiliation{Institut N\'eel, associ\'e \`a l'UJF, CNRS, BP 166,
38042 Grenoble Cedex 9, France}
\author{Theo A. Costi}
\affiliation{Institut f\"ur Festk\"orperforschung, Forschungszentrum
J\"ulich, 52425 J\"ulich, Germany}
\author{Wolfgang Wernsdorfer}
\affiliation{Institut N\'eel, associ\'e \`a l'UJF, CNRS, BP 166,
38042 Grenoble Cedex 9, France}
\author{Franck Balestro}
\affiliation{Institut N\'eel, associ\'e \`a l'UJF, CNRS, BP 166,
38042 Grenoble Cedex 9, France}

\date{\today}
\begin{abstract}
We present first quantitative experimental evidence for the underscreened Kondo
effect, an uncomplete compensation of a quantized magnetic moment by
conduction electrons, as originally proposed by Nozi\`eres and
Blandin. The device consists of an even charge spin $S=1$ molecular
quantum dot, obtained by electromigration of C$_{60}$ molecules into
gold nanogaps and operated in a dilution fridge. The persistence of
logarithmic singularities in the low temperature conductance is
demonstrated by a comparison to the fully screened configuration
obtained in odd charge spin $S=1/2$ Coulomb diamonds. We also
discover an extreme sensitivity of the underscreened Kondo resonance
to magnetic field, that we confirm on the basis of numerical
renormalization group calculations.
\end{abstract}

\maketitle

When a magnetic impurity is inserted in a piece of metal, its
magnetic moment can be completely screened by the conduction
electrons, owing to their quantized spin $1/2$. This general
phenomenon, the Kondo effect, has been thoroughly studied in diluted
magnetic alloys \cite{hewson_kondo_1993} and has attracted
considerable attention in the more recent quantum dot systems
\cite{goldhaber-gordon_kondo_1998}. Clearly, impurities carrying a
spin $S$ greater that $1/2$ need to bind several electronic orbitals
in order to fully quench their magnetism, and Nature seems to
conspire in always providing enough screening channels for that
situation to occur in general \cite{costi_kondo_2009}. Therefore,
the possibility that screening may happen to be incomplete, as
initially proposed on theoretical grounds by Nozi\`eres and Blandin
\cite{nozieres_kondo_1980}, has remained elusive for almost thirty
years, despite the great experimental control that one can achieve
with artificial quantum dot setups. The observation of the
underscreened Kondo effect, in addition to its overscreened 
counterpart \cite{potok_observation_2007}, is also especially appealing 
since it constitutes one of the simplest cases where standard Fermi 
Liquid Theory is violated \cite{Pepin,Mehta}.

We demonstrate in this Letter that molecular quantum dots obtained
through electromigration \cite{park_nanomechanical_2000} 
are perfect candidates for achieving underscreened Kondo impurities. 
Indeed point contact tunneling (single mode) and important left/right asymmetry 
of the transport electrodes ensure a large window of energies where a {\it single}
screening channel is active. In addition, the Kondo phenomenon in
molecules can set in already at several kelvins
\cite{park_coulomb_2002, liang_kondo_2002, yu_kondo_2004,
parks_tuningkondo_2007} thanks to relatively important charging
energies, allowing a complete study of Kondo crossovers on a
sufficient range of temperatures. Both conditions of single channel and
large Kondo temperature are difficult to meet altogether in other quantum 
dot devices, where Kondo effects associated with higher spin states have been 
previously found, but yet not investigated in detail \cite{schmid_absence_2000,
vanderwiel2002,kogan_singlet-triplet_2003, granger_two-stage_2005,quay_magnetic_2007}.
%
We report here on the first observation of the anomalous logarithmic behavior in
the temperature and bias voltage dependent conductance in a spin $S=1$ quantum
dot below the Kondo scale, as previously predicted for underscreened impurities
\cite{nozieres_kondo_1980,Pepin,Mehta,posazhennikova_anomalous_2005},
and successfully confront our results with quantitative numerical
renormalization group (NRG) calculations. More strikingly, the
experimental data demonstrate that underscreened impurities are
extremely sensitive to the application of a magnetic field. In that
case, and in contrast to fully screened moments, the Kondo resonance
is split by a Zeeman energy much smaller than the Kondo temperature,
reflecting the high degree of polarizability of a partially screened
spin. This surprising finding is confirmed by new NRG calculations
of the local density of states that we performed on the
single-channel spin $S=1$ Kondo model in a magnetic field.

\begin{figure}
\includegraphics[scale=0.6]{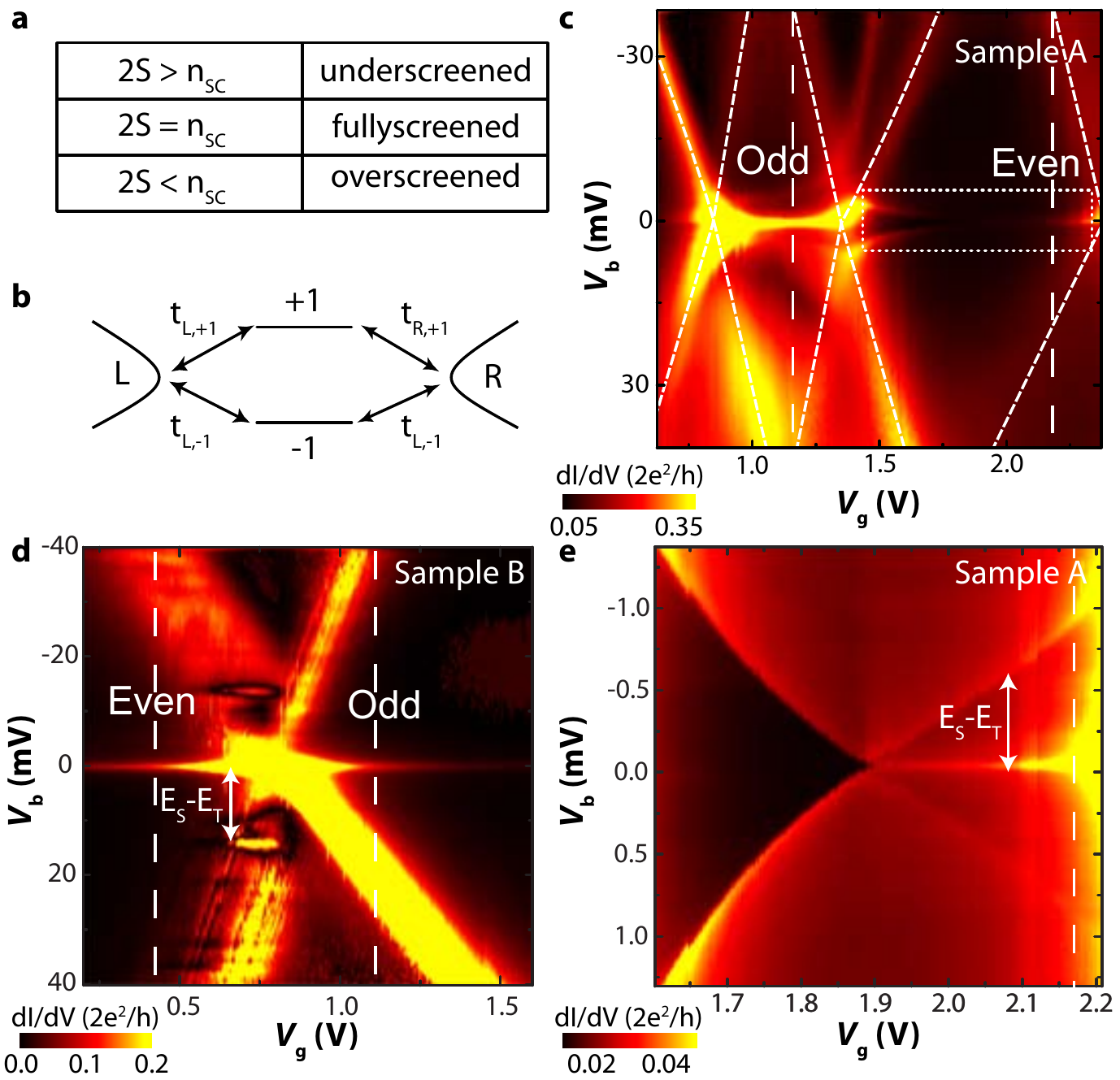}
\caption{\textbf{a.} Summary of the different types of Kondo effects
according to impurity spin $S$ and number of screening channels
$n_\mathrm{sc}$; \textbf{b.} Tunneling model of our single molecule
transistor: two orbital levels couple to two leads; \textbf{c.}
Conductance map of sample A recorded at $T= 35$mK. Dashed lines are
positioned at the gate voltages where more detailed studies where
performed; \textbf{d.} Conductance map for sample B; \textbf{e.}
Zoom inside the dotted rectangle defined in the even diamond of
sample A. The white arrows represents the singlet-triplet splitting
in both samples.} \label{figure1}
\end{figure}

Recent work by us \cite{roch_quantum_2008} has shown that C$_{60}$ 
quantum dots with two electrons on the molecule can be gate-tuned via 
a quantum phase transition between a molecular singlet and a spin $S=1$ 
Kondo state. Interestingly, the central condition for this phase transition 
to occur, vindicated by these experimental findings \cite{roch_quantum_2008}, is the
presence of a single screening channel in the accessible temperature range
\cite{pustilnik_singlet-triplet_2003,aligia2009}.
Since screening channels result from the overlap between the wave
functions of the conduction electrons and those of the magnetic
impurity orbitals~\cite{nozieres_kondo_1980}, one understands that
orbital quantum numbers must not be preserved in order for the
underscreened Kondo effect to take place, see Fig.\ref{figure1}.a
discussing the situation of $n_\mathrm{sc}$ channels coupled to an
impurity spin $S$.
We can describe our molecular transistor \cite{roch_quantum_2008} by two
orbital levels $(-1,+1)$ coupled to two metallic leads $(L,R)$ (see figure~\ref{figure1}.b),
with the tunneling matrix~\cite{pustilnik_singlet-triplet_2003}:
\begin{equation}
\label{tunnel}
\mathbf{t}=
 \begin{pmatrix}
t_{L,+1} & t_{R,+1}\\
t_{L,-1} & t_{R,-1}
\end{pmatrix}
 \end{equation}
A screening channel $\lambda = 1,2$ is associated to each eigenvalue
$t_\lambda$ of this matrix, from which result antiferromagnetic Kondo
couplings $J_\lambda=8|t_\lambda|^2/E_\mathrm{add}$ between the localized
orbitals and the conduction electrons ($E_\mathrm{add}$ is a measure of
the addition energy on the quantum dot). One then obtains two Kondo
temperatures:
\begin{equation}
\label{exp}
k_B T_{K\lambda}=\sqrt{DJ_{\lambda}}e^{-1/(\rho J_\lambda)}
\end{equation}
where $D$ is the bandwidth of the metal and $\rho$ its density of
states. While it seems physically sound that at vanishing
temperature a spin $S=1$ will be fully screened
\cite{pustilnik_singlet-triplet_2003},
moderate left/right asymmetries in the tunneling amplitudes
(\ref{tunnel}) can lead to vastly different Kondo scales, because of
the exponential behavior in (\ref{exp}), as pointed
in~\cite{posazhennikova_anomalous_2005}. This situation naturally occurs
with electromigration, as the molecule tends to stay preferably
closer to one of the electrodes, so that {\it e.g.} $t_{L,\pm1}\gg t_{R,\pm1}$.
Fig. \ref{figure1} shows indeed that our conductance maxima, here shown for two
different devices, are much lower than the quantum value $2e^2/h$, with $e$ the 
electron charge and $h$ Planck's constant. It
is then possible to have a large range of temperatures over which
underscreened behavior prevails. A further crucial condition for the
realization of the underscreened Kondo effect is the formation of a
spin $S=1$ unit itself. This relies on a strong Hund's rule that
makes the separation between singlet and triplet states much larger
than the temperature. These high-energy singlet
excitations \cite{roch_out-of-equilibrium_2008-1} are clearly
spectroscopically resolved in all our measurements through
cotunneling lines, see figure~\ref{figure1}.d-e. Sample A in
figure~\ref{figure1}.c shows an odd charge Coulomb diamond with a
pronounced Kondo ridge for gate voltages between $V_g$=0.8V and
1.35V (parity was argued in \cite{roch_quantum_2008}), so that the
next diamond seen up to $V_g$=2.3V has an even number of electrons.
Figure~\ref{figure1}.d represents the conductance map of sample B,
where a single charge degeneracy point is observed, with Kondo
ridges on {\it both} sides. We estimate a low limit for the addition
energy $E_\mathrm{add}\geqslant$200meV. Figure~\ref{figure1}.e
focuses inside the even diamond of sample A, where the singlet to triplet quantum
phase transition occurs~\cite{roch_quantum_2008}. We can thus assess
that our experiment was carried out in the temperature range
$T_{K1} \ll T \ll T_{K2}$. Similar singlet excitations are visible on the
left side of sample B, see Fig.~\ref{figure1}.d., therefore allowing
us to label this region even.

We now present the study of our two different devices (see figure~\ref{figure1}.c-d), 
both showing fully screened and underscreened Kondo anomalies. In order to maximize 
the Kondo temperature, gate voltages were chosen away from the center of the Coulomb 
diamonds,  while staying out of the mixed valence regime. In the fully
screened Kondo effect, the conductance versus lowering the temperature has a
logarithmic increase above the Kondo temperature, and then saturates in a
quadratic, Fermi-liquid like fashion. On the other hand, the underscreened
Kondo effect is expected to show two distinct logarithmic behaviors
\cite{nozieres_kondo_1980,Pepin,Mehta,posazhennikova_anomalous_2005}, 
{\it above and below} the highest Kondo temperature $T_{K2}$. 
Indeed, partially screened moments are known to act as 
a ferromagnetic Kondo impurities \cite{nozieres_kondo_1980}, leading to a 
slow logarithmic scattering of the conduction electrons at low energy.
This is best seen by plotting the derivative of conductance with respect to
temperature as a function of inverse temperature, as suggested in
\cite{posazhennikova_anomalous_2005}.
\begin{figure}
\begin{center}
\includegraphics[scale= 0.45]{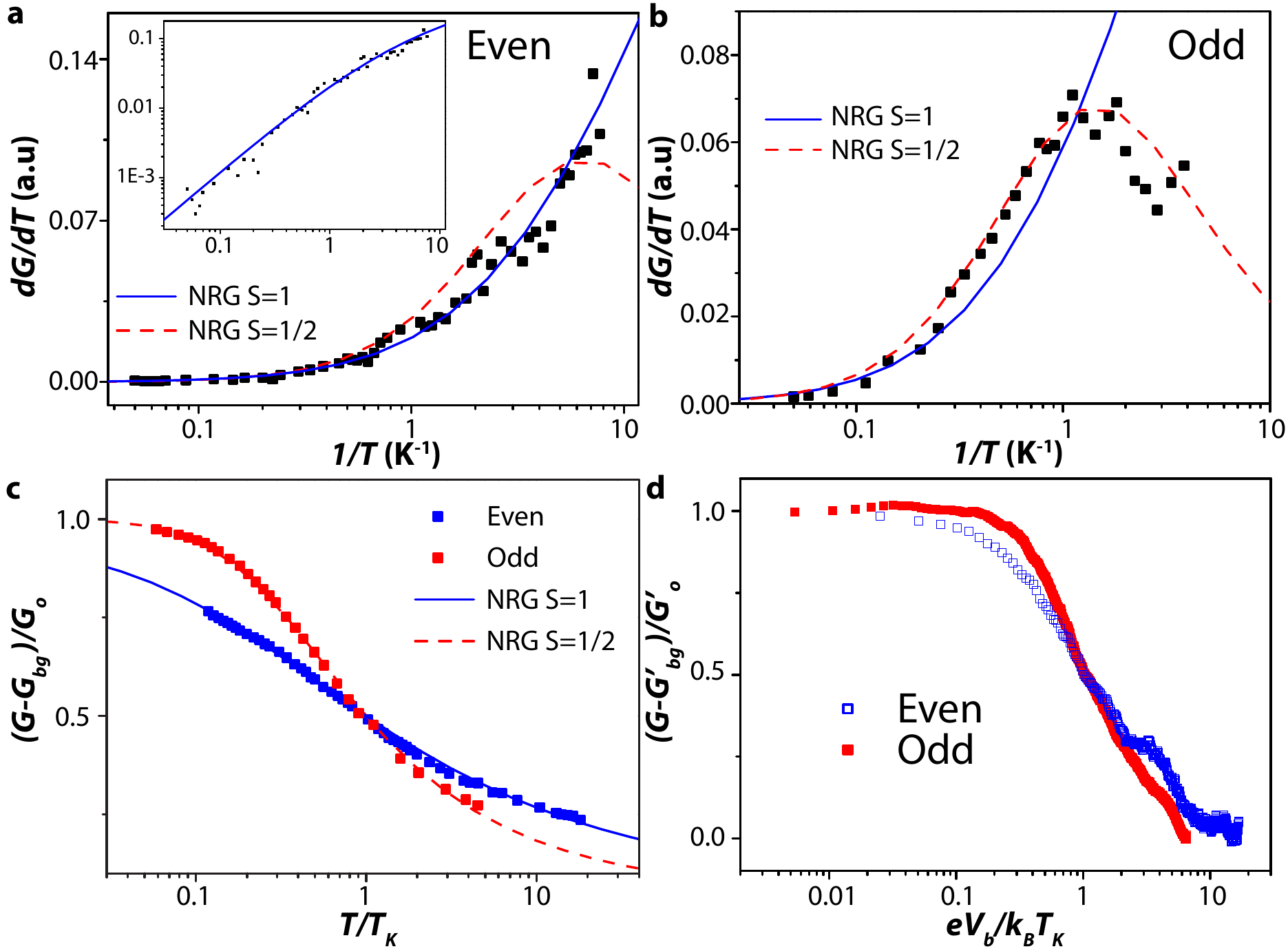}
\end{center}
\caption{\textbf{a.} Derivative of the conductance versus
temperature recorded at gate $V_g=2.17$V and bias $V_b=0$V in the even
Coulomb diamond of sample A (same data on a log scale in the inset, showing two
distinct $1/T$ behaviors). The best fit is obtained for an underscreened Kondo
model (blue) and gives $G_o=0.14$ (in units of $2e^2/h$) and $T_K=1.1$K;
\textbf{b.} Derivative of the conductance versus temperature recorded at
$V_g=1.2$V and $V_b=0$V in the odd Coulomb diamond. The best fit
is obtained for a fully screened Kondo model (red) and gives $G_o=0.34$
and $T_K=4.4$K;
\textbf{c.} Conductance data of a and b compared to the relevant NRG curves.
The background was adjusted to follow the convention $G(T_K)=G_o/2$, giving
$G_{bg}=0.055$ in the fully screened case and $G_{bg}=0.022$ in the
underscreened case;
\textbf{d.} Differential conductance at the base temperature rescaled in
universal form, with $G_o'=G(T_\mathrm{base})$, and $G_{bg}'$ such that the
conductance is $G_o'/2$ at $e V_b=k_B T_K$.} 
\label{figure2}
\end{figure}
Our data in the inset of figure~\ref{figure2}a, taken for the even diamond of
sample A, clearly display two {\it distinct} $1/T$ regimes.
Another way to discriminate both Kondo effects is to perform a fit to
NRG results~\cite{costi_kondo_2009,Mallet} of single channel spin $S=1/2$ and $S=1$
Kondo models. The obvious qualitative differences, namely the apparent
divergence in $dG(T)/dT$ at low temperature $T$ for spin $S=1$ and the
presence of a maximum for spin $S=1/2$, in addition to the quantitative
agreement with the NRG predictions are strong indications of two remarkably
different Kondo states, see Fig. \ref{figure2}a-b.
This comparison to theory allows us to extract the associated Kondo 
temperatures $T_K$ and the conductance amplitude $G_o$ for each curve, using 
the scaling form 
\begin{equation}
G(T)=G_o\,f^{\mathrm{NRG}}_S( T/T_K)+G_{bg}
\end{equation}
with $S$ the impurity spin and $G_{bg}$ a constant conductance
background~\cite{epaps}.
%
We stopped our measurements around $T_\mathrm{base}=$100mK in order to
ensure that saturation effects are not provoked by an uncontrolled
electronic temperature. We also limited the AC-excitation of the
lock-in detection to stay in the limit $eV_{AC}\ll k_B T$. With
these words of caution, we could not observe any saturation of the
conductance over two decades of temperature in the even diamond,
contrary to what is witnessed in the fully screened case. Since the
Kondo effect displays an universal behavior with a single scale
$T_K$, this must be reflected by different universal curves between
screened and underscreened situations, as evidenced by Fig.
\ref{figure2}c. The bias voltage dependence of the non-linear
conductance at the base temperature was also
examined \cite{grobis_universal_2008,scott_universal_2009}.
Although this study is more difficult to perform accurately,
two relatively different curves are nevertheless obtained, with
slower voltage behavior for the spin $S=1$ molecules, see Fig.
\ref{figure2}d. We also note that a measure of $T_K$ can be roughly 
deduced from the half width at half maximum (HWHM) of the
non-linear conductance peak, and this method was used to have
an alternative determination of $T_K$ for each Kondo ridge in sample
A and B, after extraction of a background contribution (see 
table~\ref{tableTK}).
The higher error bars for the bias method are due to the absence of
reliable finite voltage predictions from theory, so that $T_K$ 
depends sensitively on the estimated background.
Both methods, relying on thermal or voltage smearing of the Kondo
peak, nicely converge to comparable $T_K$ values for {\it both} types
of Kondo effects.
\vspace{0.5cm}
\begin{table}
\begin{center} 
\begin{tabular}{|c|c|c|c|}
                        \hline
                         Sample/Method  & Temperature & Bias & Magnetic Field\\
                        \hline
                          A (odd) & $4.4\pm 0.3K$ & $5.5\pm1.3K$ & $4.8\pm 0.3K$ \\
                        \hline
                         A (even) & $1.1\pm 0.1K$ & $1.9\pm0.5K$ & $0.6\pm 0.4K$ \\
                        \hline
                         B (odd) & none & $4.4\pm0.8K$ & $5.4\pm 0.3K$ \\
                        \hline
                         B (even) & none & $1.9\pm0.3K$ & $0.2\pm 0.1K$ \\
                        \hline
\end{tabular}
\end{center}
\caption{Kondo temperatures for each sample in the even and odd
diamonds, determined by using the methods of temperature, bias or
magnetic field, as described in the text. Sample B was not stable enough 
to perform a detailed temperature study.}
\label{tableTK}
\end{table}

\begin{figure}
\begin{center}
\includegraphics[scale= 0.47]{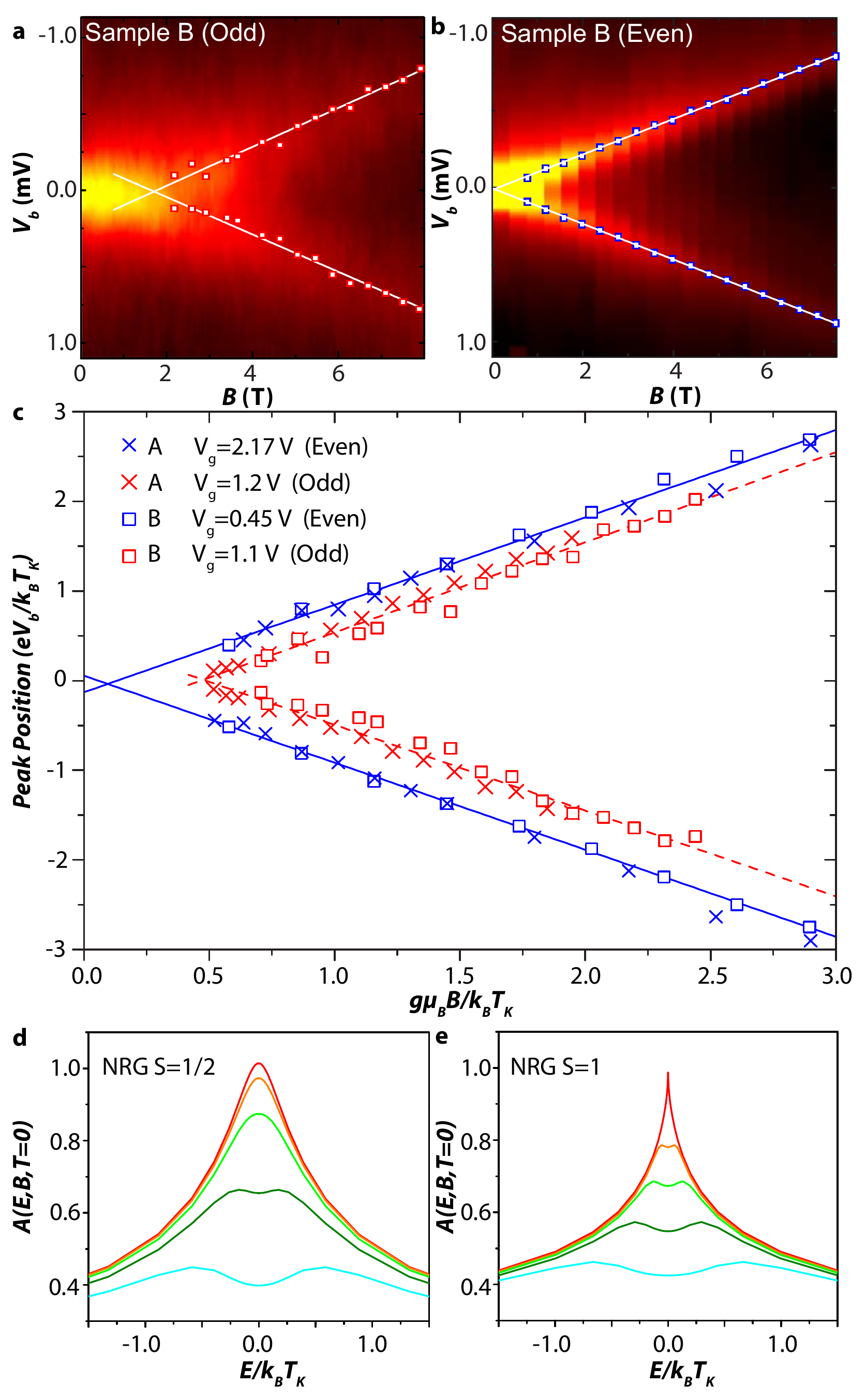}
\end{center}

\caption{\textbf{a. b.} Differential conductance versus bias voltage
and magnetic field for sample B in the odd and even diamonds (lines are 
linear fits to the B-dependence of the Zeeman peaks of sample B);
\textbf{c.} Positions of the Zeeman peaks of both samples and
odd/even diamonds, as extracted from above. Linear fits provide the 
critical field $B_c$ by extrapolation to zero bias. Both bias voltage and
magnetic field were renormalized by $T_K$ deduced from the HWHM at
$B=0$T for each case; \textbf{d. e.} NRG calculations of the energy $E$
dependent equilibrium local density of states $A(E,B,T=0)$ 
at zero temperature, normalized to its $E=B=0$ value, for the spin $S=1/2$ 
and $S=1$ single-channel Kondo models and magnetic field values 
$g \mu_B B/k_B T_K=0,1/8,1/4,1/2,1$ (top to bottom).} \label{figure3}
\end{figure}

The evolution of the Kondo peak as a function of magnetic field, as seen
in the bias spectrum of Fig. \ref{figure3}.a-b, displays even more dramatic 
effects.
Previous experimental studies
\cite{cronenwett_tunable_1998,kogan_measurements_2004,quay_magnetic_2007}
and theoretical calculations \cite{costi_kondo_2000} on fully
screened impurities have shown a Zeeman splitting of the Kondo
resonance at a critical field $g \mu_B B_c\sim 0.5 k_B  T_K$. Indeed, as seen in
table~\ref{tableTK}, this estimate agrees with the $T_K$ determined by the 
temperature and bias dependence, for the odd charge diamonds of both samples. 
In contrast, {\it for the spin $S=1$ quantum dot, the Kondo resonance separates into Zeeman 
states at a much lower magnetic field}, $g \mu_B B_c \ll k_B T_K$, so 
that the Kondo temperature naively determined from $2g\mu_B B_c$ is in 
disagreement with the previous estimates, see table~\ref{tableTK}.
This observation can be understood as the combination of two related
effects. First, a partially screened Kondo impurity is expected to
be highly polarizable in a small magnetic field, which constitutes a
relevant perturbation to the free spin fixed point, in the renormalization 
group sense. Second, the logarithmic cusp of the
Kondo peak at low bias makes the splitting more pronounced than for
the Lorentzian shape typical of a Fermi Liquid in the fully screened
case. To put this surprising observation on firmer ground, we have
performed NRG calculations for the local density of states in a
magnetic field for the single-channel Kondo models with spins
$S=1/2$ and $S=1$. This numerical solution allows us to confirm that
the underscreened Kondo resonance starts to unveil its magnetic
excitations for Zeeman energies as low as  $k_B T_K/16$, see figure
\ref{figure3}.d-e. A further interesting check on our analysis lies
on the extracted Zeeman splitting as a function of magnetic field,
figure \ref{figure3}.c. For odd/even diamonds respectively, we find
that the dispersion of the magnetic states, rescaled to the
corresponding Kondo temperature, clearly shows the presence/absence
of a threshold. Also remarkable is how these data for our two {\it
different} samples quite naturally fall on top of each other,
another signature of Kondo universality.

In conclusion, we have given comprehensive experimental evidence for
the occurence of Nozi\`eres and Blandin underscreened Kondo effect in even
charge molecular quantum dots. Our analysis, based on the stark differences with
respect to regular transport properties of fully screened impurities, was 
strenghtened by NRG calculations. An unexpected 
magnetic field sensitivity of partially screened Kondo impurities was also 
discovered, that we could confirm theoretically. This work illustrates the striking 
versatility of molecular electronics for the investigation of fundamental aspects in 
quantum magnetism.

We thank E. Eyraud, D. Lepoittevin for their technical
contributions, E. Bonet, T. Crozes, T. Fournier for lithography
development, and C. Winkelmann, C. Thirion, M. Desmukh, L. Calvet
for useful discussions. Samples were fabricated in the NANOFAB
facility of the N\'eel Institute. This work is partially financed by
ANR-PNANO projects  MolSpintronics No. ANR-06-NANO-27,  MolNanoSpin
n°ANR-08-NANO-002, ERC Advanced Grant MolNanoSpin n°226558, and STEP
MolSpinQIP.

\bibliographystyle{apsrev}

\newpage
\cleardoublepage
\setcounter{figure}{0}
\setcounter{equation}{0}

\section{Supplementary Information on ``Observation of the underscreened 
Kondo effect in a molecular transistor''}

\subsection{On the Kondo scaling of the temperature-dependent conductance}

A hallmark of Kondo dominated quantum transport in nanostructures is
the universal one-parameter scaling of the linear conductance as a function of
the ratio of temperature $T$ to the Kondo scale $T_K$:
\begin{equation}
G(T) = G_o\; f(T/T_K)+G_{bg}
\end{equation}
with $G_0$ the overall magnitude of the conductance variation (related
to the asymmetry in lead couplings and the total hybridization $\Gamma$ 
to the leads), $G_{bg}$ a high
temperature background associated with direct tunneling processes, and
$f(x)$ an universal scaling function that depends solely on the spin $S$ carried
by the quantum dot.
An important feature of this physical quantity is the broad crossover
taking place between the two extreme limits of temperature, $T\ll T_K$ and $T\gg T_K$.
For the spin $S=1/2$ fully screened case, the scaling function behaves as 
$f(x)=1-(\pi^4/16)x^2$ and $f(x)=(3\pi^2/16)/\log^2(x)$ in the Fermi
liquid ($x\ll1$) and local moment ($x\gg1$) regimes respectively
\cite{Hewson}. The complete 
crossover curve, associated with the full function $f(x)$ for all $x$, can be 
accurately calculated from Wilson's NRG \cite{Costi}

\begin{figure*}
\includegraphics[scale=1.0]{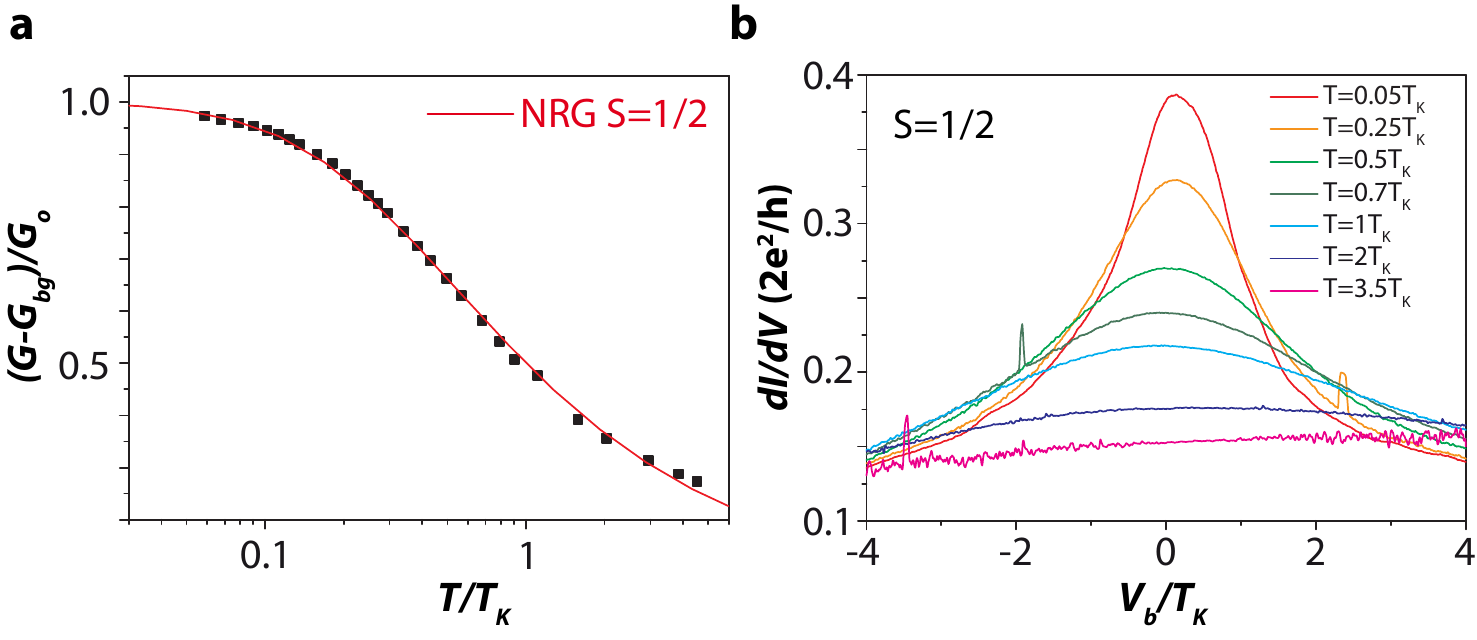}
\caption{{\bf a.} Linear conductance as a function of temperature for sample A in the 
half-integer spin region (gate voltage $V_g=1.2$V), rescaled in universal form, and
compared to NRG calculations on the spin $S=1/2$ Kondo model. 
{\bf b.} Related non-linear conductance traces as a function of bias (rescaled by the
Kondo scale $T_K$) for several different temperatures.}
\label{fig1}
\end{figure*}

In practice, however, the universal high temperature regime is 
attained in experiment only if the Kondo scale $T_K$ happens to 
be at least two orders of magnitude smaller than the 
Coulomb repulsion $U$ on the quantum dot \cite{Oguri}.
In that case, Coulomb blockade effects
have already set in at temperatures much higher than $T_K$, ensuring that one can 
safely extract $f(x)$ for $x\gg1$, corresponding to the logarithmic Kondo tails. 
In terms of the microscopic Anderson model, this condition
corresponds to $U/\pi\Gamma > 2$, where the universal crossover from the local moment 
to the Fermi liquid regimes takes place (the Wilson ratio is then quite close to 
$2$) \cite{Oguri}. This clearly sets a first important constraint on the value 
of the charging energy when analyzing an experiment in terms of universal Kondo behavior.
Now, to assess experimentally the low temperatures regime, associated with the
behavior of $f(x)$ at $x\ll1$, 
it is crucial that the actual Kondo temperature is not vanishingly small, preferably 
an order of magnitude higher than the base temperature $T_\mathrm{base}$ of the cryostat. 
When the above two conditions, $10T_\mathrm{base}<T_K< U/100$, are met together, 
the conductance shows a characteristic inverted-S shape, see 
Fig.\ref{fig1}a, that provides a very strong constraint to the fitting 
procedure with the NRG calculation.
In this case, all non-universal parameters
($T_K$, $G_o$, $G_{bg}$) can be accurately extracted from the data. On the
contrary, when the Kondo scale does not obey one of these two requirements, the 
incomplete determination of the crossover function makes the fit quite sensitive 
to the estimation of the background $G_{bg}$, resulting in an imprecise 
determination of $T_K$.
In our experiment the Coulomb energy $U$ is of the order of $1000$K, the 
Kondo scale $T_K$ is $4$K (for the spin $S=1/2$ samples), and our temperature 
ranges from $100$mK to $20$K, so that we can clearly capture the {\it whole} Kondo 
crossover, and determine $T_K$ unambiguously. The full development of the Kondo
resonance is similarly observed in the non-linear conductance, Fig.\ref{fig1}b,
which becomes effectively very broad in the high temperature regime, while
saturating to a constant height and width below $200$mK.

\begin{figure}
\includegraphics[scale=0.95]{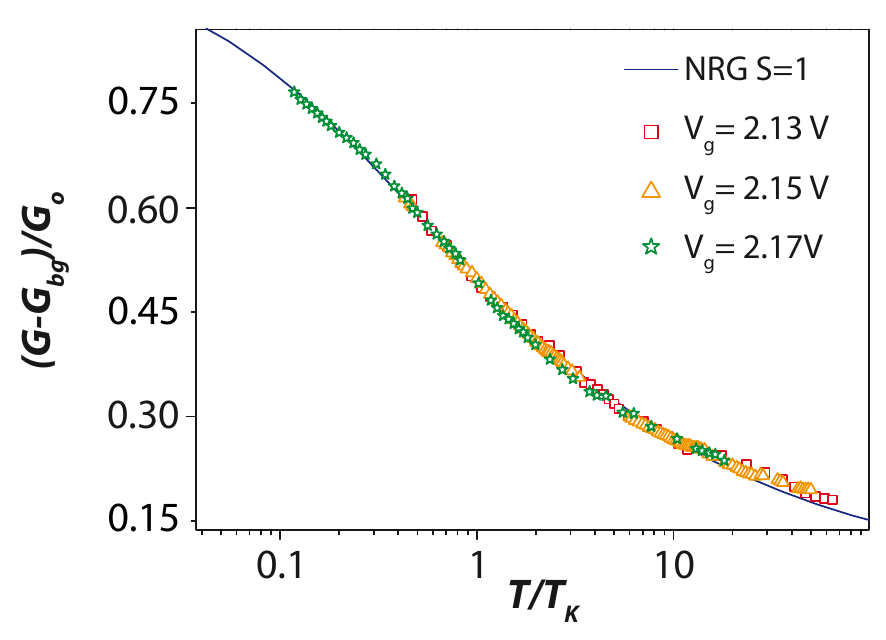}
\caption{Linear conductance as a function of temperature for sample A in the integer 
spin region (gate voltages $V_g=2.13, 2.15, 2.17$V), rescaled in universal form, and 
compared to NRG calculations on the spin $S=1$ Kondo model.} 
\label{fig2}
\end{figure}

The method to address the underscreened Kondo effect in our integer spin samples
follows exactly the same strategy of fitting the complete crossover function to
the universal result from NRG calculations on the spin $S=1$ Kondo model.
Here, the Kondo scale $T_K$ is found to be close to $1K$ at the gate voltage 
$V_g=2.17V$ for sample A, using a quantitative fit to the NRG result as shown 
in Fig.\ref{fig2}.
Again, this conductance displays both the Kondo tails (for $T>T_K$) and a
slightly different behavior (for $T<T_K$). The latter clearly appears to be much slower 
than the clear saturation measured in the spin $S=1/2$ samples (see Fig.\ref{fig1}a) and is 
associated with the second logarithmic regime {\em below} the Kondo scale, which is 
a characteristic ``smoking gun'' signature of the underscreened Kondo 
effect. This interpretation is further strengthened by the quantitative agreement
to the NRG calculations.
Measurements at two further gate voltages have also been taken, 
and data collapse onto the universal curve for the 
underscreened Kondo conductance is reasonably good, see Fig.\ref{fig2}. We stress, 
however, that these two extra measurements, performed nearer to the center of the 
Coulomb diamond, are associated with relatively smaller Kondo temperatures in the
$200$mK range, and do not allow to probe the {\em whole} Kondo crossover from
well below to well above $T_{K}$, in contrast to our measurement at the optimal
gate voltage at 2.17V. The key to 
maximizing $T_K$ is thus to benefit from the virtual charge fluctuations on the 
side of the Coulomb diamond, while staying away from the mixed valence regime, 
leading to an optimal choice of voltage where the experiment should be performed.

\subsection{Zeeman effect: threshold behavior in the non-linear conductance}

\begin{figure}
\includegraphics[scale=0.55]{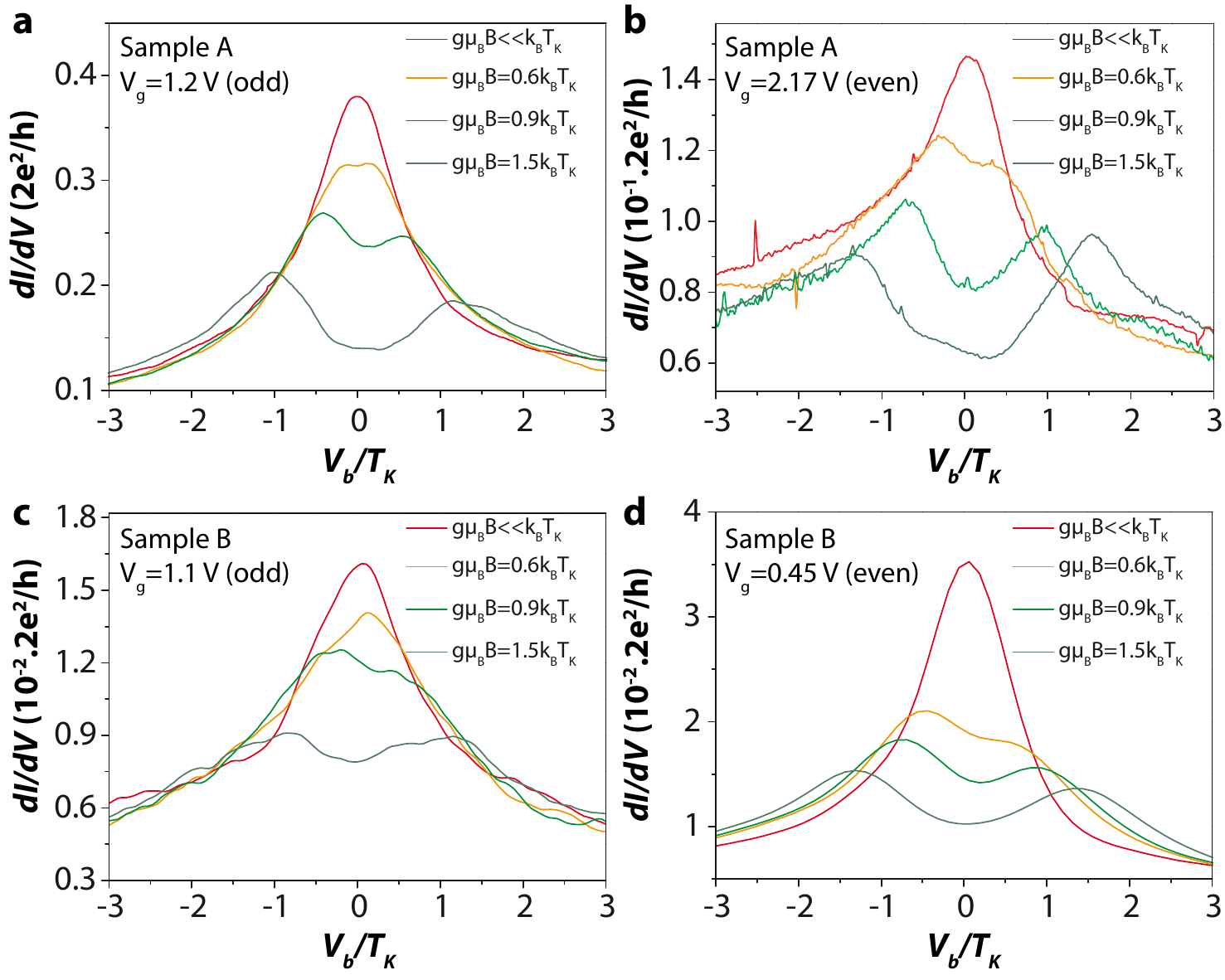}
\caption{Non-linear conductance traces as a function of bias (rescaled by the
Kondo scale $T_K$) for fixed values of the ratio of magnetic field $B$ to the Kondo
temperature $T_K$. Regions associated to spin $S=1/2$ and $S=1$ for sample A, taken at 
the base temperature, are shown in panels {\bf a} and {\bf b} respectively, which 
corresponds to the linear conductances given in Figs.\ref{fig1}a and \ref{fig2}. 
Similar data for sample B are shown in panels {\bf c} and {\bf d}, corresponding 
to odd and even Coulomb diamonds respectively.}
\label{fig3}
\end{figure}

A second key signature of the underscreened Kondo effect (with respect to the standard 
fully screened case) lies in an anomalous Zeeman effect of the Kondo resonance,
as found in our study.
For completeness, we present here the raw non-linear conductance traces, that 
display the same data, yet in a different fashion than the color plots given in 
the main text.  Here again, odd and even charge diamonds for two different samples
are investigated at ``optimal'' gate voltages (see discussion above), and
the conductance traces at the base temperature for several values of
the magnetic field are shown in Fig.\ref{fig3}. 
All data are rescaled by the corresponding Kondo temperature $T_K$, both 
with respect to bias voltage $V_b$ and magnetic field $B$. It is clearly 
apparent that the Zeeman splitting in the Kondo peak occurs at lower 
values of $B/T_K$ in the case of spin $S=1$ samples. This phenomenon reflects the 
high degree of polarizability of the underscreened Kondo impurity, and this 
anomalous Zeeman effect constitutes a crucial fingerprint of the 
underscreened Kondo effect, that can be used as a guide for further experimental 
investigations.

\end{document}